# "Relativistic" particle dynamics without relativity

Allan Walstad*
*Physics Department, University of Pittsburgh at Johnstown, Johnstown, PA 15904 USA*

Abstract: It is shown that the correct expressions for momentum and kinetic energy of a particle moving at high speed were already implicit in physics going back to Maxwell. The demonstration begins with a thought experiment of Einstein by which he derived the inertial equivalence of energy, independently of the relativity postulates. A simple modification of the same experiment does the rest.

## I. INTRODUCTION

In classical electromagnetic theory, a directed pulse of light with energy E has momentum[1] $p = E/c$. Using this relation and very little else, Einstein[2] demonstrated with a thought experiment the mass equivalence (inertia) of energy via $E = mc^2$. Although coming after his first papers on relativity, this particular argument was entirely independent of the relativity postulates. It does not appear to have been noticed that with Einstein's result in hand a simple modification of the same thought experiment is sufficient to establish the formulas for momentum and energy of a material particle traveling at high speed. One therefore arrives at "relativistic" particle dynamics without benefit of the special theory of relativity.

## II. EINSTEIN'S BOX

French[3] provides a textbook discussion of Einstein's thought experiment. Consider a box of length *L* and mass *M*, initially at rest, as shown in Fig. 1.

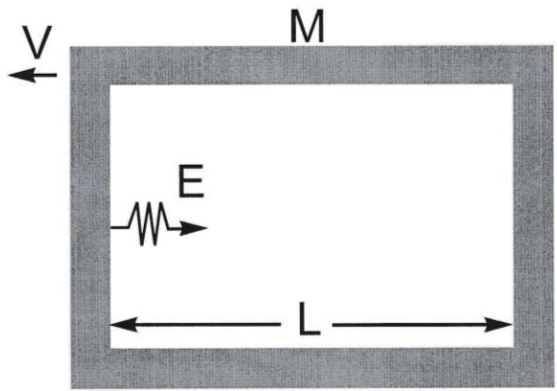

Fig. 1: Einstein's box. A pulse of light having energy E is emitted from one end of a box of length L and mass M. As a result, the box recoils at speed V.

A pulse of light having energy *E* is emitted from the left end and absorbed at the right. When the light is emitted, the box must recoil with momentum $p = E/c$, hence with speed

$$V = \frac{E}{Mc} \tag{1}$$

(which we are taking to be slow enough that there is no question about the accuracy of the Newtonian expression for momentum). The box is brought to rest when the light is absorbed. Nevertheless, during the light travel time $t = L/c$, the box moved to the left a distance

$$\delta s = Vt = \frac{EL}{Mc^2}. \tag{2}$$

Thus it would appear that the center of mass of an isolated, initially stationary system has shifted spontaneously. This result can be avoided if the energy transferred is associated with a mass $\delta m = E/c^2$. In that case we will have $M\,\delta s = L\,\delta m$, which is the condition that the center of mass not move.[4]

The problem to be avoided by the assignment of mass to energy is more than just a small one-time shift in center of mass of the box. In Einstein's discussion (with further elaboration by Max Von Laue[5]), if the energy can be returned by means other than light back to the left side, the process can be repeated indefinitely, violating conservation of momentum. Rather than speculating on alternative means of energy transfer, it may be conceptually more straightforward to imagine the box equipped with a gyroscopic wheel by which it can rotate about its center of mass to reverse the ends; then the energy can be returned again via a pulse of light, and with the ends reversed the center of mass would be translated farther in the same direction.

### III. THE MODIFICATION

We consider now an apparently novel extension of the thought experiment in which the light pulse is replaced by a material particle of (rest) mass m, emitted at (high) speed $v$ with momentum $p$ and kinetic energy $K$. At the outset, we do not know the relationship among these quantities. The kinetic energy, when deposited in the right end of the box and dissipated as heat, cannot be distinguished from the electromagnetic energy that was deposited in the earlier version. The total mass transferred from left to right, then, must be $m + K/c^2$. The recoiling box, with mass $M - m - K/c^2$, travels at speed

$$V = \frac{p}{M - m - \frac{K}{c^2}} \tag{3}$$

for time $t = L/(v + V)$, a distance $\delta s = Vt$ to the left. Meanwhile, the mass carried with the particle travels a distance $L - \delta s$ to the right. The condition for stationary center of mass is

$$\left(M - m - \frac{K}{c^2}\right) \delta s = \left(m + \frac{K}{c^2}\right)(L - \delta s). \tag{4}$$

Substituting for $\delta s$ in terms of the other parameters and solving for $p$, we find

$$p = (m + K/c^2)v. \tag{5}$$

As a particle is accelerated from rest, the work done by the net force goes into kinetic energy, so we have

$$dK = F\, dz \tag{6a}$$

$$= (dp/dt)(v\, dt) \tag{6b}$$

$$= v\, dp. \tag{6c}$$

We can eliminate the momentum using Eq. (5), then integrate up to find $K = m(\gamma - 1)c^2$, where $\gamma = (1 - v^2/c^2)^{-1/2}$ is the familiar Lorentz factor. Alternatively, we can eliminate $K$ and find $p = m\gamma v$. These results establish what we know as relativistic particle dynamics.

It should be noted that French[6] also obtains the relativistic expressions for momentum and kinetic energy prior to taking up the relativity postulates, but only by speculating that one can combine $E = mc^2$ with the Newtonian expression $p = mv$ to obtain $E = c^2 p/v$. Davidon[7] also arrived at the relativistic expressions by means of similar assumptions added to the inertial equivalence of energy. The modified thought experiment described in this paper gives these results a secure basis.

## IV. WAS EINSTEIN'S ARGUMENT VALID?

Einstein's thought experiment generated a small literature of tweaks aimed at simplicity and improved rigor. See Antippa[8] for a review. We review here an objection raised by Feenberg,[9] as it has been some time since the issue was aired. Einstein took the box to be rigid, so that when the light pulse is emitted the entire box recoils at once. Of course, the recoil cannot be transmitted along the sides of the box faster than the speed of sound in the material, which is much slower than the speed of light. Therefore, the sequence of events is other than was assumed: the left end recoils to the left when the light is emitted, the right end recoils to the right when the light is absorbed, and only much later are the two ends halted by tension in the sides of the box.

French[10] addresses this concern by eliminating the sides of the box, leaving just the ends. Let each end have mass $M/2$. If the light pulse of energy $E$ does not carry mass, then the left end recoils to the left with speed

$$V = \frac{2E}{Mc} \tag{7}$$

and, at time $t = L/c$ later, the right end recoils to the right with equal speed. Once both ends are moving, the center of mass remains fixed, but since the left end started first, the center of mass has been shifted to the left, and it is easy to see that the distance is

$$\delta s = \frac{EL}{Mc^2} \tag{8}$$

as for the rigid box. If desired, we can arrange for both ends to be brought to rest by means of a long, light thread that initially has considerable slack. As the ends move apart, the thread gradually becomes taught, bringing them to rest, after which it can be used further to pull them to their original separation, all of this taking place symmetrically so as not to disturb the now-shifted center of mass. Then we are left with an isolated system that began at rest and ended at rest, for which the sole change has been a translation of the center of mass.

Suppose instead that the emitted light pulse carries mass $\delta m$. The left end, now with mass $M/2 - \delta m$, recoils left with speed

$$V_L = \frac{E}{\left(\frac{M}{2} - \delta m\right)c} \tag{9}$$

After time $t = L/c$, when the light has reached the right end, the condition that the center of mass of the system has not moved is

$$\left(\frac{M}{2} - \delta m\right) V_L \, t = L \, \delta m \, . \tag{10}$$

The condition is satisfied with $\delta m = E/c^2$. After the light is absorbed at the right end, the two ends move apart with equal and opposite momenta; hence, the center of mass remains fixed. Einstein's result, as presented in section II, is thereby confirmed.

*awalstad@pitt.edu*

[1] Feynman gave an extremely concise pedagogical derivation of the momentum of light: Richard P. Feynman, Robert Leighton, and Matthew Sands, *The Feynman Lectures on Physics*, Vol. 1 (Addison-Wesley, Reading, MA, 1963), Ch. 34, pp. 10-11.

[2] Albert Einstein, "The principle of conservation of motion of the center of gravity and the inertia of energy," Annalen der Physik (ser. 4) **20**, 627-633 (1906). English translation in *The Collected Papers of Albert Einstein, The Swiss Years: Writings, 1900-1909*, Vol. 2, translated by Anna Beck (Princeton Univ. Press, Princeton, NJ, 1989), pp. 200-206.

[3] A. P. French, *Special Relativity* (Norton, New York, 1968), pp. 16-29.

[4] More rigorously, the recoil speed is

$$V = \frac{E}{(M-\delta m)c},$$

the time is $t = L/(c + V)$, and the condition of stationary center of mass is

$$(M - \delta m)\delta s = \delta m(L - Vt).$$

After some algebra the result $\delta m = E/c^2$ is unchanged.